\documentclass[12pt,a4paper]{article}
\usepackage{graphicx}

\usepackage{subeqnarray,indent,amsfonts,amssymb}
\usepackage{eqsection}
\usepackage{bm}    
\usepackage{cite}  
\usepackage{graphicx}

\catcode`\@=11
\def\marginnote#1{}

\relax

\def\beq{\begin{equation}}
\def\eeq{\end{equation}}
\def\bea{\begin{eqnarray}}
\def\eea{\end{eqnarray}}
\def\beaa{\begin{array}}
\def\eeaa{\end{array}}

\begin{document}

\title{\vspace*{-30mm}\hspace*{-13mm}Eigenvalue amplitudes of the Potts model on a torus\hspace*{-13mm}}

\author{
  {\small Jean-Fran\c{c}ois Richard${}^{1,2}$ and
          Jesper Lykke Jacobsen${}^{1,3}$} \\[1mm]
  {\small\it ${}^1$Laboratoire de Physique Th\'eorique
  et Mod\`eles Statistiques}                             \\[-0.2cm]
  {\small\it Universit\'e Paris-Sud, B\^at.~100,
             91405 Orsay, France}                        \\[1mm]
  {\small\it ${}^2$Laboratoire de Physique Th\'eorique
  et Hautes Energies}                                    \\[-0.2cm]
  {\small\it Universit\'e Paris VI,
             Bo\^{\i}te 126, Tour 24, 5${}^{\mbox{\`eme}}$ {\'e}tage} \\[-0.2cm]
  {\small\it 4 place Jussieu, 75252 Paris cedex 05, France} \\[1mm]
  {\small\it ${}^3$Service de Physique Th\'eorique}      \\[-0.2cm]
  {\small\it CEA Saclay, Orme des Merisiers,
             91191 Gif-sur-Yvette, France}               \\[-0.2cm]
  {\protect\makebox[5in]{\quad}}  
  \\
}

\maketitle
\thispagestyle{empty}   

\begin{abstract}

  We consider the $Q$-state Potts model in the random-cluster
  formulation, defined on {\em finite} two-dimensional lattices of size $L
  \times N$ with toroidal boundary conditions. Due to the non-locality
  of the clusters, the partition function $Z(L,N)$ cannot be written
  simply as a trace of the transfer matrix ${\rm T}_L$. Using a combinatorial
  method, we establish the decomposition $Z(L,N) = \sum_{l,D_k}
  b^{(l,D_k)} K_{l,D_k}$, where the characters $K_{l,D_k} = \sum_i
  (\lambda_i)^N$ are simple traces. In this decomposition, the amplitudes
  $b^{(l,D_k)}$ of the eigenvalues $\lambda_i$ of ${\rm T}_L$ are labelled
  by the number $l=0,1,\ldots,L$ of clusters
  which are non-contractible with respect to the transfer ($N$) direction,
  and a representation $D_k$ of the cyclic group $C_l$. We obtain rigorously a
  general expression for $b^{(l,D_k)}$ in terms of the characters of $C_l$,
  and, using number theoretic results, show that it coincides with an
  expression previously obtained in the continuum limit by Read and Saleur.

\end{abstract}


\section{Introduction}

The $Q$-state Potts model on a graph $G=(V,E)$ with vertices $V$ and edges $E$
can be defined geometrically through the cluster expansion of the partition
function \cite{FK}
\begin{equation}
 Z=\sum_{E' \subseteq E} Q^{n(E')} ({\rm e}^J-1)^{b(E')} \,,
 \label{Zcluster}
\end{equation}
where $n(E')$ and $b(E')=|E'|$ are respectively the number of connected
components (clusters) and the cardinality (number of links) of the edge
subsets $E'$. We are interested in the case where $G$ is a finite regular
two-dimensional lattice of width $L$ and length $N$, so that $Z$ can be
constructed by a transfer matrix ${\rm T}_L$ propagating in the $N$-direction.

In \cite{cyclic}, we studied the case of cyclic boundary
conditions (periodic in the $N$-direction and non-periodic in the
$L$-direction). We decomposed $Z$ into linear combinations of certain
restricted partition functions (characters) $K_l$ (with $l=0,1,\ldots,L$) in
which $l$ {\em bridges} (that is, marked non-contractible clusters) wound
around the transfer ($N$) direction. We shall often refer to $l$ as the {\em
level}. Unlike $Z$ itself, the $K_l$ could be written as (restricted) traces
of the transfer matrix, and hence be directly related to its eigenvalues. It
was thus straightforward to deduce from this decomposition the amplitudes in
$Z$ of the eigenvalues of ${\rm T}_L$.
The goal of this work is to repeat this procedure in the
case of toroidal boundary conditions. 

Note that as in the cyclic case some other procedures exist. First,
Read and Saleur have given in~\cite{read} a general formula for the
amplitudes, based on the earlier Coulomb gas analysis of Di Francesco,
Saleur, and Zuber~\cite{zuber}.  They obtained that the amplitudes of
the eigenvalues are simply $b^{(0)}=1$ at the level $l=0$ and
$b^{(1)}=Q-1$ at $l=1$.  For $l\geq 2$ they obtained that, contrary to
the cyclic case, there are several differents amplitudes at each level $l$.
Their number is equal to $q(l)$, the number of divisors of $l$. They
are given by:
\begin{equation}
b^{(l,m)}=\Lambda(l,m;e_0)+(Q-1)\Lambda \left(l,m;\frac{1}{2} \right)\;, 
\label{eqsal}
\end{equation}
where $l$ is the level considered, and $m$ is a divisor of $l$ which
labels the different amplitudes for a given level.
$\Lambda$ is defined as:

\begin{equation}
\Lambda(l,m;e_0)=2 \sum_{d>0 \, : \, d|l} \frac{\mu\left( \frac{m}{m\wedge d}\right)\phi\left( \frac{l}{d} \right)}
{l \, \phi\left( \frac{m}{m\wedge d} \right)} \cos(2\pi d e_0) \;.
\label{defLambda}
\end{equation}
Here, $\mu$ and $\phi$ are respectively the M\"obius and Euler's totient function~\cite{hardy}. The 
M\"obius function $\mu$ is defined by $\mu(n)=(-1)^r$, if $n$ is an integer that is a product 
$n=\prod_{i=1}^r p_i$ of $r$ {\em distinct} primes, $\mu(1)=1$, and $\mu(x)=0$ otherwise or if $x$
is not an integer. Similarly, Euler's totient function $\phi$ is defined for positive integers $n$
as the number of integers $n'$ such that $1\leq n'\leq n$ and $n\wedge n'=1$.
The value of $e_0$ depends on $Q$ and is given by:
\begin{equation}
\sqrt{Q}=2\cos (\pi e_0 )
\label{defe0}
\end{equation}
Note that in Eq.~(\ref{defLambda}) we may write
$\cos(2\pi d e_0) = T_{2d}(\sqrt{Q}/2)$, where $T_n(x)$ is the $n$'th order
Chebyshev polynomial of the first kind.
The term $(Q-1)\Lambda(l,m;\frac{1}{2})$ in Eq.~(\ref{eqsal}) is due to configurations containing a cluster with ``cross-topology'' \cite{zuber,read} (see later).

The drawback of the derivation in Ref.~\cite{read} is that since it relies
ultimately on free-field techniques it is {\em a priori} valid only at the
usual ferromagnetic critical point ($J=J_{\rm c}$) and in the continuum limit
($N,L\to\infty$). But one may suspect, in analogy with the cyclic case, that
these amplitudes would be valid for any finite lattice and for any
inhomogeneous (i.e., edge-dependent) values of the coupling constants $J$.

To our knowledge, no algebraic study proving this statement does
exist in the literature. Indeed, when the boundary conditions
are toroidal, the transfer matrix (of the related six-vertex model, to
be precise) does no longer commute with the generators of the quantum
group $U_q(sl(2))$. Therefore, there is no simple algebraic way of
obtaining the amplitudes of eigenvalues, although some progress has
been made by considering representations of the periodic
Temperley-Lieb algebra. A good review is given by Nichols
\cite{nichols}.

Chang and Shrock have studied the Potts model with
toroidal conditions from a combinatorial point of view \cite{shrock}.
Using a diagrammatic approach they obtained some general results on the
eigenvalue amplitudes. In particular, they showed that the sum of all
amplitudes at level $l$ equals
\begin{equation}
 b^{(l)} \equiv \sum_{j=0}^l b_j^{(l)}
 = \left \lbrace \begin{array}{ll}
 \sum_{j=0}^l (-1)^{l-j} \frac{2l}{l+j} {l+j \choose  l-j} Q^j + (-1)^l (Q-1)
 & \mbox{for }l \ge 2 \\
 \sum_{j=0}^l (-1)^{l-j} {l+j \choose l-j } Q^j
 & \mbox{for }l \le 2 \\
 \end{array} \right.
\label{defbltori}
\end{equation}
They also argued that it was because ${\rm T}_L$ enables permutations among the  bridges, due to
the periodic boundary conditions in the transverse ($L$) direction, that there were different amplitudes for a given 
level $l$. Without them, all the amplitudes at level $l$ would be equal (to a global factor) to $b^{(l)}$.
Finally, they computed explicitly the amplitudes at levels $l=2$ and $l=3$; one may check that those results are in agreement with Eq.~(\ref{eqsal}).

Using the combinatorial approach we developed in~\cite{cyclic}, we will make
the statements of Chang and Shrock more precise, and we will give in
particular a new interpretation of the amplitudes using the characters of the
cyclic group $C_l$. Then, by calculating sums of characters of irreducible
representations (irreps) of this group, we will reobtain Eq.~(\ref{eqsal}) and
thus prove its validity for an arbitrary finite $L \times N$ lattice. As will
become clear below, the argument relies exclusively on counting correctly the
number of clusters with non-trivial homotopy, and so the conclusion will hold
true for any edge-dependent choice of the coupling constants $J$ as well.

Our approach will have to deal with several complications due to the boundary
conditions, the first of which is that the bridges can now be permuted (by
exploiting the periodic $L$-direction). In the following this leads us to
consider decomposition of $Z$ into more elementary quantities than $K_l$,
namely characters $K_{l,P}$ labeled by $l$ {\em and} a permutation of the
cyclic group $C_l$. However, $K_{l,P}$ is not simply linked to the eigenvalues
of $T$, and thus we will further consider its expansion over related
quantities $K_{l,D_k}$, where $D_k$ labels an irreducible representation
(irrep) of $C_l$. It is $K_{l,D_k}$ which are the elementary quantities in the
case of toroidal boundary conditions.%
\footnote{In a previous publication on the same subject \cite{torus1} 
we have studied the decomposition in terms of the full symmetric group
$S_l$. The present approach, using only the cyclic group $C_l$, is far
simpler and for the first time allows us to prove Eq.~(\ref{eqsal}).
Note also that some misprints had cropped up in Ref.~\cite{torus1},
giving in particular wrong results for the amplitudes at level $l=4$.}

The structure of the article is as follows. In section \ref{sec2}, we define
appropriate generalisations of the quantities we used in the cyclic case
\cite{cyclic} and we expose all the mathematical background we will need.
Then, in section \ref{sec3}, we decompose restricted partition functions---and
as a byproduct the total partition function---into characters $K_l$ and
$K_{l,P}$. Finally, in section \ref{sec4}, we obtain a general expression of
the amplitudes of eigenvalues which involves characters of irreps of $C_l$.
Using number theoretic results (Ramanujan sums) we then proceed to prove its
equivalence with the formula~(\ref{eqsal}) of Read and Saleur.

\section{Algebraic preliminaries}
\label{sec2}

\subsection{Definition of the $Z_{j,n1,P}$}

As in the cyclic case, the existence of a periodic boundary condition allows
for non-trivial clusters (henceforth abbreviated NTC), i.e., clusters which
are not homotopic to a point. However, the fact that the torus has {\em two}
periodic directions means that the topology of the NTC is more complicated
that in the cyclic case. Indeed, each NTC belongs to a given homotopy class,
which can be characterised by two coprime numbers $(n_1,n_2)$, where $n_1$
(resp.\ $n_2$) denotes the number of times the cluster percolates horizontally
(resp.\ vertically) \cite{zuber}. The fact that all clusters (non-trivial or
not) are still constrained by planarity to be non-intersecting induces a
convenient simplification: all NTC in a given configuration belong to the same
homotopy class. For comparison, we recall that in the cyclic case the only
possible homotopy class for a NTC was $(n_1,n_2)=(1,0)$.

It is a well-known fact \cite{pasquier,RJ2} that the difficulty in decomposing
the Potts model partition function---or relating it to partition functions of
locally equivalent models (of the six-vertex or RSOS type)---is due solely to
the weighing of the NTC. Although a typical cluster configuration will of
course contain trivial clusters (i.e., clusters that are homotopic to a point)
with seemingly complicated topologies (e.g., trivial clusters can surround
other trivial clusters, or be surrounded by trivial clusters or by NTC), we
shall therefore tacitly disregard such clusters in most of the arguments that
follow. Note also that a NTC that span both lattice directions%
\footnote{Such a cluster was referred to as ``degenerate'' in Ref.~\cite{RJ2},
and as a cluster having ``cross-topology'' in Ref.~\cite{zuber}.}
in the present context corresponds to $n_1=1$.

\begin{figure}
  \centering
  \includegraphics[width=150pt]{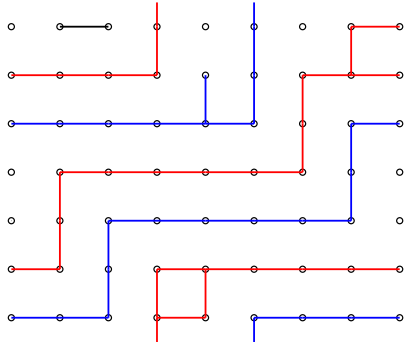}
  \caption{Cluster configuration with $j=2$ non-trivial clusters (NTC),
 here represented in red and blue colours. Each NTC is characterised by
 its number of branches, $n_1=2$, and by the permutation it realises,
 $P=(12)$. Within a given configuration, all NTC have the same topology.}
  \label{fig1}
\end{figure}

Consider therefore first the case of a configuration having a single NTC. For
the purpose of studying its topology, we can imagine that is has been shrunk
to a line that winds the two periodic directions $(n_1,n_2)$ times. In our
approach we focus on the the properties of the NTC along the direction of
propagation of the transfer matrix ${\rm T}_L$, henceforth taken as the
horizontal direction. If we imagine cutting the lattice along a vertical line,
the NTC will be cut into $n_1$ horizontally percolating parts, which we shall
call the $n_1$ {\em branches} of the NTC. Seen horizontally, a given NTC
realises a permutation $P$ between the vertical coordinates of its $n_1$
branches, as shown in Fig.~\ref{fig1}. Up to a trivial relabelling of the
vertical coordinate, the permutation $P$ is independent of the horizontal
coordinate of the (imaginary) vertical cut, and so, forms part of the
topological description of the NTC. We thus describe totally the topology
along the horizontal direction of a NTC by $n_1$ and the permutation $P \in
S_{n_1}$.

Note that there are restrictions on the admissible permutations $P$. Firstly,
$P$ cannot have any proper invariant subspace, or else the corresponding NTC
would in fact correspond to several distinct NTC, each having a smaller value
of $n_1$. For example, the case $n_1=4$ and $P=(13)(24)$ is not admissible, as
$P$ corresponds in fact to two distinct NTC with $n_1=2$. In general,
therefore, the admissible permutations $P$ for a given $n_1$ are simply cyclic
permutations of $n_1$ coordinates. Secondly, planarity implies that the
different branches of a NTC cannot intersect, and so not all cyclic
permutations are admissible $P$. For example, the case $n_1=4$ and $P=(1324)$
is not admissible. In general the admissible cyclic permutations are
characterised by having a constant coordinate difference between two
consecutive branches, i.e., they are of the form $(k,2k,3k,\ldots)$ for some
constant $k$, with all coordinates considered modulo $n_1$. For example, for
$n_1=4$, the only admissible permutations are then finally $(1234)$ and
$(1432)$.%
\footnote{Note that we consider here the permutations that can be realised by
a {\em single} cluster, not all the admissible permutations at a given level.
We shall come back to this issue later (in Sec.~\ref{sec:decKlCl}) when we
discuss in detail the attribution of ``black points'' to one or more different
NTC. It will then be shown that the admissible permutations at level $l$
correspond to the cyclic group $C_l$. For example, the admissible permutations
at level $l=4$ are ${\rm Id}$, $(1234)$, $(13)(24)$ and $(1432)$.}

Consider now the case of a configuration with several NTC. Recalling that all
NTC belong to the same homotopy class, they must all be characterised by the
same $n_1$ and $P$. Alternatively one can say that the branches of the
different NTC are entangled. Henceforth we denote by $j$ the number of NTC
with $n_1\geq 1$ in a given configuration. Note in particular that, seen along
the horizontal direction, configurations with no NTC and configurations with
one or more NTC percolating only vertically are topologically equivalent.
This is an important limitation of our approach.

Let us denote by $Z_{j,n_1,P}$ the partition function of the Potts model on an
$L \times N$ torus, restricted to configurations with exactly $j$ NTC
characterised by the index $n_1\geq 1$ and the permutation $P \in S_{n_1}$; if
$P$ is not admissible, or if $n_1 j > L$, we set $Z_{j,n_1,P}=0$. Further, let
$Z_{j,n_1}$ be the partition function restricted to configurations with $j$
NTC of index $n_1$, let $Z_j$ be the partition function restricted to
configurations with $j$ NTC {\em percolating horizontally}, and let $Z$ be
the total partition function. Obviously, we have $Z_{j,n_1}=\sum_{P\in
S_{n_1}} Z_{j,n_1,P}$, and $Z_{j}=\sum_{n_1=1}^{L} Z_{j,n_1}$, and
$Z=\sum_{j=0}^{L} Z_{j}$. In particular, $Z_0$ corresponds to the partition
function restricted to configurations with no NTC, or with NTC percolating
only vertically.

In the case of a generic lattice all the $Z_{j,n_1,P}$ are non-zero, provided
that $P$ is an admissible cyclic permutation of length $n_1$, and that $n_1 j
\leq L$. The triangular lattice is a simple example of a generic lattice. Note
however that other regular lattices may be unable to realise certain
admissible $P$. For example, in the case of a square lattice or a honeycomb
lattice, all $Z_{j,n_1,P}$ with $n_1 j =L$ and $n_1 > 1$ are zero, since there
is not enough ``space'' on the lattice to permit all NTC branches to percolate
horizontally while realising a non-trivial permutation. Such non-generic
lattices introduce additional difficulties in the analysis which have to be
considered on a case-to-case basis. In the following, 
we consider therefore the case of a generic lattice.

\subsection{Structure of the transfer matrix}

The construction and structure of the transfer matrix ${\rm T}$ can be taken
over from the cyclic case \cite{cyclic}. In particular, we recall that ${\rm
T}$ acts towards the right on states of connectivities between two time slices
(left and right) and has a block-trigonal structure with respect to the number
of {\em bridges} (connectivity components linking left and right) and a
block-diagonal structure with respect to the residual connectivity among the
non-bridged points on the left time slice. As before, we denote by ${\rm T}_l$
the diagonal block with a fixed number of bridges $l$ and a trivial residual
connectivity. Each eigenvalue of ${\rm T}$ is also an eigenvalue of one or
more ${\rm T}_l$. In analogy with \cite{shrock} we shall sometimes call ${\rm
T}_l$ the transfer matrix at level $l$. It acts on connectivity states which
can be represented graphically as a partition of the $L$ points in the right
time slice with a special marking (represented as a {\em black point}) of
precisely $l$ distinct components of the partition (i.e., the components that
are linked to the left time slice via a bridge).

A crucial difference with the cyclic case is that for a given partition of the
right time slice, there are more possibilities for attributing the black
points (for $0 < l < L$). Considering for the moment the black points
to be indistinguishable, we denote the corresponding dimension as $n_{\rm
tor}(L,l)$. It can be shown \cite{shrock} that
\beq
 n_{\rm tor}(L,l) = \left \lbrace
 \begin{array}{ll}
   \frac{1}{L+1} {2L \choose L} & \mbox{for }l=0 \\
   {2L-1 \choose L-1}           & \mbox{for }l=1 \\
   {2L \choose L-l}             & \mbox{for }2 \le l \le L
 \end{array} 
 \right.
 \label{defntor}
\eeq
and clearly $n_{\rm tor}(L,l)=0$ for $l>L$.

Suppose now that a connectivity state at level $l$ is time evolved by a
cluster configuration of index $n_1$ and corresponding to a permutation $P$.
This can be represented graphically by adjoining the initial connectivity
state to the left rim of the cluster configuration, as represented in
Fig.~\ref{fig1}, and reading off the final connectivity state as seen from the
right rim of the cluster configuration. Evidently, the positions of the black
points in the final state will be permuted with respect to their positions in
the intial state, according to the permutation $P$. As we have seen, not all
$P$ are admissible. We will show in the subsection~\ref{sec:decKlCl} that the
possible permutations at a given level $l$ (taking into account all the ways of
attributing $l$ black points to cluster configurations) are the elements of the cyclic group $C_l$.%
\footnote{We proceed differently from Chang and Shrock~\cite{shrock} who
considered the group $S_l$ of all permutations at level $l$, not just the
admissible permutations. Therefore the dimension of ${\rm T}_l$ they obtained
was $l! \; n_{\rm tor}(L,l)$. Although this approach is permissible (since in
any case ${\rm T}_l$ will have zero matrix elements between states which are
related by a non-admissible permutation) it is more complicated \cite{torus1}
than the one we present here.}
the number of possible connectivity states without taking into account the
possible permutations between black points, the dimension of $T_l$ is $l \;
n_{\rm tor}(L,l)$, as $C_l$ has $l$ distinct elements.

Let us denote by $|v_{l,i} \rangle$ (where $1\leq i \leq n_{\rm tor}(L,l)$)
the $n_{\rm tor}(L,l)$ standard connectivity states at level $l$. The full
space of connectivities at level $l$, i.e., with $l$ distinguishable black
points, can then be obtained by subjecting the $|v_{l,i} \rangle$ to
permutations of the black points. It is obvious that ${\rm T}_l$ commutes with
the permutations between black points (the physical reason being that ${\rm
T}_l$ cannot ``see'' to which positions on the left time slice each bridge is
attached). Therefore ${\rm T}_l$ itself has a block structure in a appropriate
basis. Indeed, ${\rm T}_l$ can be decomposed into ${\rm T}_{l,D}$ where ${\rm
T}_{l,D}$ is the restriction of ${\rm T}_l$ to the states transforming
according to the irreducible representation (irrep) $D$ of $C_l$. 
Note that as $C_l$ is a abelian group of $l$ elements, it has $l$ irreps of 
dimension $1$.
One can obtain the corresponding basis by applying the
projectors $p_D$ on all the connectivity states at level $l$, where $p_D$ is
given by
\begin{equation}
 p_D=\frac{1}{l} \sum_{P} \bar{\chi}_D(P) \, P \;.
\label{projpD}
\end{equation}
Here $\chi_D(P)$ is the
character of $P$ in the irrep $D$ and $\bar{\chi}_D(P)$ is its complex conjugate. 
The application of all permutations of $C_l$ on any given
standard vector $|v_{l,i} \rangle$ generates a regular representation of
$C_l$, which contains therefore once each representation $D$ (of dimension $1$).
As there are $n_{\rm tor}(L,l)$ standard vectors,
the dimension of ${\rm T}_{l,D}$ is thus simply $n_{\rm
tor}(L,l)$.%
\footnote{Note that if had considered the group $S_l$ instead of $C_l$ we
would have had algebraic degeneracies, which would have complicated
considerably the determination of the amplitudes of eigenvalues. In fact, it
turns out that even by considering $C_l$ there are degeneracies between
eigenvalues of different levels, as noticed by Chang and Shrock~\cite{shrock}.
But these degeneracies depend of the width $L$, and have no simple algebraic
interpretation.}

\subsection{Definition of the $K_{l,D}$}
\label{sec:defKlD}

We now define, as in the cyclic case \cite{cyclic}, $K_l$ as the trace of
$\left({\rm T}_l\right)^N$. Since ${\rm T}_l$ commutes with $C_l$, we can
write
\begin{equation}
K_l=l \sum_{i=1}^{n_{\rm tor}(L,l)} \langle v_{l,i}| \left({\rm T}_l\right)^N
|v_{l,i} \rangle \; .
\label{defKltor}
\end{equation}
In distinction with the cyclic case, we cannot decompose the partition
function $Z$ over $K_l$ because of the possible permutations of black points
(see below). We shall therefore resort to more elementary quantities, the
$K_{l,D}$, which we define as the trace of $\left({\rm T}_{l,D}\right)^N$.
Since both ${\rm T}_l$ and the projectors $p_D$ commute with $C_l$, we have
\begin{equation}
K_{l,D}=l \sum_{i=1}^{n_{\rm tor}(L,l)} \langle v_{l,i}| p_D \left({\rm T}_l\right)^N |v_{l,i} \rangle \; .
\label{defKlDtor}
\end{equation}
Obviously one has
\begin{equation}
K_l = \sum_D K_{l,D} \;,
\end{equation}
the sum being over all the $l$ irreps $D$ of $C_l$. Recall that in the cyclic case
the amplitudes of the eigenvalues at level $l$ are all identical. This is no
longer the case, since the amplitudes depend on $D$ as well. Indeed
\begin{equation}
K_{l,D}=\sum_{k=1}^{n_{\rm tor}(L,l)} 
\left(\lambda_{l,D,k}\right)^N \; .
\label{defastrace}
\end{equation}

In order to decompose $Z$ over $K_{l,D}$ we will first 
use auxiliary quantities, the $K_{l,P_l}$ defined as:
\begin{equation}
K_{l,P_l}=\sum_{i=1}^{n_{\rm tor}(L,l)} \langle v_{l,i}|\left(P_l\right)^{-1} \left(T_l\right)^N |v_{l,i} \rangle \; ,
\label{defKlPtor}
\end{equation}
$P_l$ being an element of the cyclic group $C_l$. So $K_{l,P_l}$ can be
thought of as modified traces in which the final state differs from the
initial state by the application of $P_l$. Note that $K_{l,{\rm Id}}$ is
simply equal to $\frac{K_l}{l}$. Because of the possible permutations of the
black points, the decomposition of $Z$ will contain not only the $K_{l,{\rm
Id}}$ but also all the other $K_{l,P_l}$, with $P_l \in C_l$. We will show
that the coefficients before $K_{l,P_l}$ coincide for all $P_l \in C_l$
that belong to the same class {\em with respect to the symmetric group $S_l$}.%
\footnote{Since $C_l$ is an abelian group, each of its elements defines
a class of its own, if the notion of class is taken with respect to $C_l$
itself. What we need here is the non-trivial classes defined with respect
to $S_l$.}
We will note these classes $(d_i,n_1)$ (corresponding to a level $l=d_i n_1$)
and it is thus natural to define $K_{(d_i,n_1)}$ as:
\begin{equation}
K_{(d_i,n_1)}=\sum_{P_l \in (d_i,n_1)} K_{l,P_l} \;,
\label{defKlCtor}
\end{equation}
the sum being over elements $P_l \in C_l$ belonging to the class $(d_i,n_1)$. 
This definition will enable us to simplify some formulas, but ultimately we will come back to the $K_{l,P_l}$.

Once we will obtain the decomposition of $Z$ into $K_{l,P_l}$, we will need to express 
the $K_{l,P_l}$ in terms of the $K_{l,D}$ to obtain the decomposition of $Z$ into $K_{l,D}$, which
are the quantities directly linked to the eigenvalues.
Eqs.~(\ref{defKlDtor}) and
(\ref{projpD}) yield a relation between $K_{l,D}$ and $K_{l,P_l}$:
\begin{equation}
K_{l,D}= \sum_{P_l} \chi_D(P_l) K_{l,P_l} \; .
\label{KlDfC}
\end{equation}
These relations can be inverted so as to obtain $K_{l,P_l}$ in terms of
$K_{l,D}$, since the number of elements of $C_l$ equals the number of irreps
$D$ of $C_l$. Multiplying Eq.~(\ref{KlDfC}) by $\bar{\chi}_D(P'_l)$ and
summing over $D$, and using the orthogonality relation $\sum_D
\bar{\chi}_D(P_l) \chi_D(P'_l) = l \delta_{P_l,P'_l}$ one easily deduces that:
\begin{equation}
K_{l,P_l}=\sum_D \frac{\bar{\chi}_D(P_l)}{l} K_{l,D}
\label{KlCfD}
\end{equation}
Note that
\begin{equation}
 \sum_D \bar{\chi}_D(P_l) = l \, \delta_{P_l,{\rm Id}}
 \label{cDClsum}
 \label{relcDC}
\end{equation}

\subsection{Useful properties of the group $C_l$}
\label{Clprop}

In the following we will obtain an expression of the amplitudes at the level
$l$ which involves sums of characters of the irreps $D$ of $C_l$. In order to
reobtain Eq.~(\ref{eqsal}), we will have to calculate these sums. We give here
the results we shall need.

$C_l$ is the group generated by the permutation $E_l=(12\dots l)$. It is
abelian and consists of the $l$ elements $E_l^a = (E_l)^a$, with $1\leq a \leq
l$.%
\footnote{With the chosen convention, the identity corresponds to $a=l$.}
The cycle structure of these elements is given by a simple rule.
We denote by $d_i$ (with $1\leq i \leq q(l)$) the integer divisors of $l$
(in particular $d_1=1$ and $d_{q(l)}=l$), 
and by $A_{d_i}$ the set of integers which are a product of $d_i$ by an
integer $n$ such that $1\leq n \leq \frac{l}{d_i}$ and
$n\wedge \frac{l}{d_i}=1$,%
\footnote{Note that the union of all the sets $A_{d_i}$ is $\{ 1,2,\dots,l\}$.}
If $a\in A_{d_i}$ then $E_l^a$ consists of $d_i$ entangled cycles of the same
length $\frac{l}{d_i}$. We denote the corresponding class $\left( d_i,
\frac{l}{d_i}\right)$. The number of elements of $A_{l_i}$, and so the number
of such $E_l^a$, is equal to $\phi \left( \frac{l}{d_i} \right)$, where $\phi$
is Euler's totient function whose definition has been recalled in the
introduction.%
\footnote{Note that $\sum_{d_i|l} \phi \left( \frac{l}{d_i} \right)=l$.}

Consider $C_6$ as an example. The elements of $C_6$ in the class $(1,6)$ are
$E_6=(123456)$ and $E_6^5=(165432)$. The elements in $(2,3)$ are
$E_6^2=(135)(246)$ and $E_6^4=(153)(264)$.%
\footnote{Note that for example $(123)(456)$ is not an element of $C_6$
since it is not entangled.}
There is only one element $E_6^3=(14)(25)(36)$ in $(3,2)$, and only
$E_6^6={\rm Id}$ in $(6,1)$. Indeed, the integer divisors of $6$ are $1$, $2$,
$3$, $6$, and we have $A_1=\{ 1,5 \}$, $A_2=\{ 2,4 \}$, $A_3=\{ 3 \}$, $A_6=\{
6 \}$.

$C_l$ has $l$ irreps denoted $D_k$, with $1 \leq k \leq l$. The corresponding
characters are given by $\chi_{D_k}\left( E_l^a\right)=\exp \left( -i 2\pi
\frac{k a}{l} \right)$.%
\footnote{With the chosen convention, 
the identity representation is denoted $D_l$.}
We will have to calculate in the following the sums given by:
\begin{equation}
\sum_{P_l \in \left(d_i,\frac{l}{d_i}\right)} \bar{\chi}_{D_k}(P_l)=\sum_{a\in A_{d_i}}  
\exp \left( i 2\pi \frac{k a}{l} \right) \; .
\end{equation}
These sums are slight generalizations of Ramanujan's sums.%
\footnote{The case where the sum is over $a\in A_{1}$ corresponds
exactly to a Ramanujan's sum.}
Using Theorem $272$ of Ref.~\cite{hardy}, we obtain that:
\begin{equation}
\sum_{P_l \in \left(d_i,\frac{l}{d_i}\right)} \bar{\chi}_{D_k}(P_l)=
\frac{\mu\left( \frac{m}{m\wedge d_i} \right) \phi \left( \frac{l}{d_i} \right)} 
{\phi \left( \frac{m}{m\wedge d_i} \right)} \; ,
\label{passage}
\end{equation}
where $k$ is supposed to be in $A_{d}$ and $m$ is given by $\frac{l}{d}$. The
M\"obius function $\mu$ has been defined in the Introduction. Note that all
$k$ which are in the same $A_{d}$ lead to the same sum; we can therefore
restrain ourselves to $k$ equal to an integer divisor of $l$ in order to have
the different values of these sums. Indeed, we will label the different
amplitudes at level $l$ by $m$.

\section{Decomposition of the partition function}
\label{sec3}

\subsection{The characters $K_l$}
\label{sec:charKl}

\begin{figure}
  \centering
  \includegraphics[width=300pt]{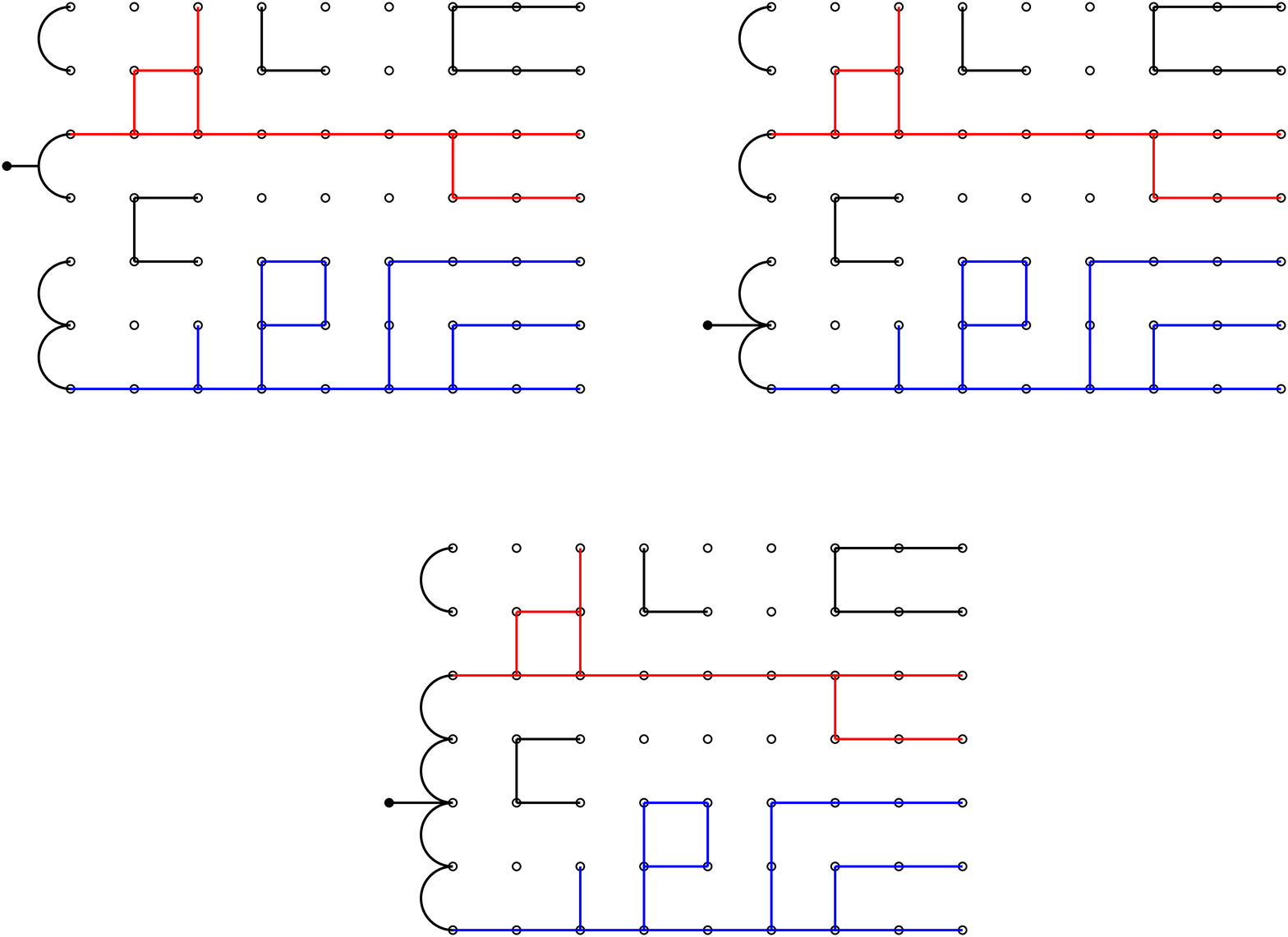}
  \caption{Standard connectivity states at level $l=1$ which are compatible
 with a given cluster configuration contributing to $Z_{2,1}$.}
  \label{fig2}
\end{figure}

By generalising the working for the cyclic case, we can now obtain a
decomposition of the $K_l$ in terms of the $Z_{j,n_1}$. To that end, we first
determine the number of states $|v_{l,i}\rangle$ which are {\em compatible}
with a given configuration of $Z_{j,n_1}$, i.e., the number of initial states
$|v_{l,i}\rangle$ which are thus that the action by the given configuration
produces an identical final state. The notion of compatibility is illustrated
in Fig.~\ref{fig2}.

We consider first the case $n_1=1$ and suppose that the $k$'th NTC connects
onto the points $\{y_k\}$. The rules for constructing the compatible
$|v_{l,i}\rangle$ are identical to those of the cyclic case:
\begin{enumerate}
 \item The points $y \notin \cup_{k=1}^j \{ y_k \}$ must be connected
 in the same way in $|v_{l,i}\rangle$ as in the cluster configuration.
 \item The points $\{y_k\}$ within the same bridge must be connected
 in $|v_{l,i}\rangle$.
 \item One can independently choose to associate or not a black point to
 each of the sets $\{y_k\}$. One is free to connect or not two distinct sets
 $\{y_k\}$ and $\{y_{k'}\}$.
\end{enumerate}
The choices mentioned in rule 3 leave $n_{\rm tor}(j,l)$ possibilities for
constructing a compatible $|v_{l,i}\rangle$. The coefficient of $Z_{j,1}$ in
the decomposition of $K_l$ is therefore $\frac{l \; n_{\rm tor}(j,l)}{Q^j}$,
since the allowed permutation of black points in a standard vector $|v_{l,i}\rangle$
allows for the construction of $l$ distinct states, and since the weight of
the $j$ NTC in $K_l$ is $1$ instead of $Q^j$. It follows that
\beq
 K_l = \sum_{j=l}^L l \, n_{\rm tor}(j,l) \frac{Z_{j,1}}{Q^j} \qquad
 \mbox{for } n_1=1.
\eeq

\begin{figure}
  \centering
  \includegraphics[width=300pt]{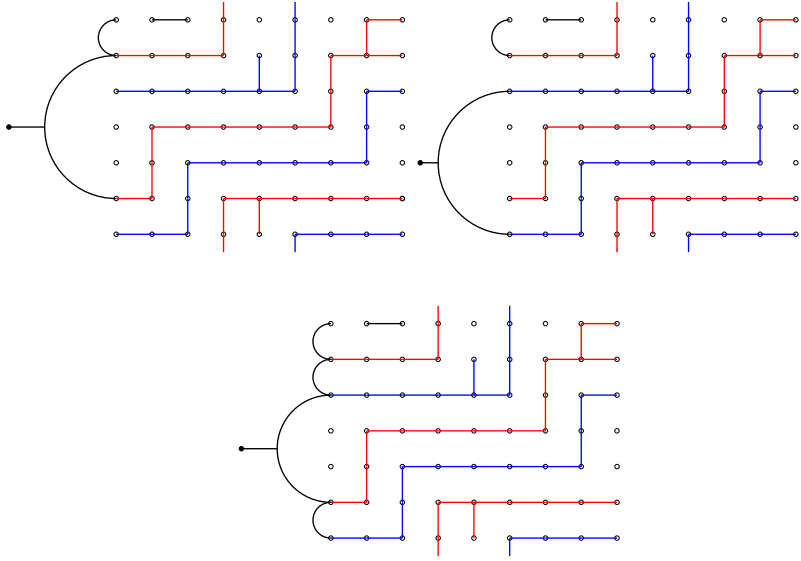}
  \caption{Standard connectivity states at level $l=1$ which are compatible
 with a given cluster configuration contributing to $Z_{2,2}$.}
  \label{fig3}
\end{figure}

We next consider the case $n_1>1$. Let us denote by $\{y_{k,m}\}$ the points
that connect onto the $m$'th branch of the $k$'th NTC (with $1 \le m \le n_1$
and $1 \le k \le j$), and by $\{y_k\}=\cup_{m=1}^{n_1}\{y_{k,m}\}$ all the
points that connect onto the $k$'th NTC.
As shown in Fig.~\ref{fig3}, the $|v_{l,i}\rangle$ which are compatible
with this configuration are such that
\begin{enumerate}
 \item The connectivities of the points $y\notin\cup_{k=1}^j\{y_k\}$ are
 identical to those appearing in the cluster configuration.
 \item All points $\{y_{k,m}\}$ corresponding to the branch of a NTC must
 be connected.
 \item We must now count the number of ways we can link the branches of the $k$ NTC and attribute
 $l$ black points so that the connection {\em and the position of the black points}
 are unchanged after action of the cluster configuration. For $l\geq 2$, there are 
 no compatible states (indeed it is not possible to respect planarity and
 to leave the position of the black points unchanged). For $l=1$ and $l=0$ there are 
 respectively ${2j-1 \choose j}=\frac12 \cdot {2j \choose j}$ and ${2j \choose j}$ 
 compatible states. Note that these results do not depend on the precise
 value of $n_1$ (for $n_1>1$).
\end{enumerate}
The rule $3$ implies that the decomposition of $K_l$ with $l\geq 2$ does not contain
any of the $Z_{j,n_1}$ with $n_1>1$. We therefore have simply
\begin{equation}
 K_l=\sum_{j=l}^L l \, n_{\rm tor}(j,l) \frac{Z_{j,1}}{Q^j} \qquad
 \mbox{for }l \geq 2 \;.
\label{expKltor}
\end{equation}
The decomposition of $K_1$ and $K_0$ are given by:
\begin{equation}
K_1=\sum_{j=1}^L n_{\rm tor}(j,1)\frac{Z_{j,1}}{Q^j} + \sum_{j=1}^{\left
\lfloor \frac{L}{2}\right \rfloor} \frac{{2j \choose j}}{2} \frac{Z_{j,n_1>1}}{Q^j}
\label{expK1tor}
\end{equation}
\begin{equation}
K_0=\sum_{j=0}^L n_{\rm tor}(j,0)\frac{Z_{j,1}}{Q^j} + \sum_{j=1}^{\left\lfloor\frac{L}{2}\right\rfloor} {2j \choose j}\,
\frac{Z_{j,n_1>1}}{Q^j} \; .
\label{expK0tor}
\end{equation}
Note that the coefficients in front of $Z_{j,n_1}$ do not depend on the
precise value of $n_1$ when $n_1>1$. To simplify the notation we have defined
$Z_{0,1}=Z_0$.

\subsection{The coefficients $b^{(l)}$}
\label{sec:coef_bl}

Since the coefficients in front of $Z_{j,1}$ and $Z_{j,n_1>1}$ in
Eqs.~(\ref{expK1tor})--(\ref{expK0tor}) are different, we cannot
invert the system of relations (\ref{expKltor})--(\ref{expK0tor})
so as to obtain $Z_j \equiv Z_{j,1}+Z_{j,n_1>1}$ in terms of the $K_l$.
It is thus precisely because of NTC with several branches contributing to
$Z_{j,n_1>1}$ that the problem is more complicated than in the cyclic case.

In order to appreciate this effect, and compare with the precise results
that we shall find later, let us for a moment assume that Eq.~(\ref{expKltor})
were valid also for $l=0,1$. We would then obtain
\begin{equation}
Z_{j,1}=\sum_{l=j}^L b_j^{(l)} \frac{K_l}{l} \; ,
\label{expZj1i}
\end{equation}
where the coefficients $b^{(l)}_j$ have already been defined in
Eq.~(\ref{defbltori}).
The coefficients $b^{(l)}$ play a role analogous to those denoted $c^{(l)}$ in
the cyclic case \cite{cyclic}; note also that $b^{(l)}= c^{(l)}$ for $l \leq
2$. Chang and Schrock have developed a diagrammatic technique for obtaining
the $b^{(l)}$ \cite{shrock}.

Supposing still the unconditional validity of Eq.~(\ref{expKltor}), one
would obtain for the full partition function
\begin{equation}
Z=\sum_{l=0}^L b^{(l)} \frac{K_l}{l} \; .
\label{devZtorosim}
\end{equation}
This relation will be modified due to the terms $Z_{j,n_1>1}$ realising
permutations of the black points, which we have here disregarded. To get
things right we shall introduce irrep dependent coefficients
$b^{(l,D)}$ and write $Z=\sum_{l=0}^L \sum_D b^{(l,D)}K_{l,D}$. 
Neglecting $Z_{j,n_1>1}$ terms would lead, according to
Eq.~(\ref{devZtorosim}), to $b^{(l,D)}=\frac{b^{(l)}}{l}$ independently of
$D$. We shall see that the $Z_{j,n_1>1}$ will lift this degeneracy of
amplitudes in a particular way, since there exist certain relations between
the $b^{(l,D)}$ and the $b^{(l)}$.

In order to simplify the formulas we will obtain later, we define the coefficients
$\tilde{b}^{(l)}$ for $l\geq 1$ by:
\begin{equation} 
\tilde{b}^{(l)} = 
 \sum_{j=0}^l (-1)^{l-j} \frac{2l}{l+j} {l+j \choose  l-j} Q^j + (-1)^l (Q-1)\; .
\label{defbltil}
\end{equation}
For $l\geq 2$, $\tilde{b}^{(l)}$ is simply equal to $b^{(l)}$, they are different  
only for $l=1$, as we have $b^{(1)}=Q-1$ but $\tilde{b}^{(1)}=-1$.  
In order to reobtain the expression (\ref{eqsal}) of Read and Saleur
for the amplitudes we will use that:
\begin{equation}
\tilde{b}^{(l)} =
2\cos (2\pi l e_0) + (-1)^l (Q-1)\; ,
\label{def2bltil}
\end{equation}
where $e_0$ has been defined in Eq.~(\ref{defe0}).

\subsection{Decomposition of the $K_{l,P_l}$}
\label{sec:decKlCl}

The relations~(\ref{expKltor})--(\ref{expK0tor}) were not invertible due to an
insufficient number of elementary quantities $K_l$. Let us now show how to
produce a development in terms of $K_{l,P_l}$, i.e., taking into account the
possible permutations of black points. This development turns out to be
invertible.

A standard connectivity state with $l$ black points is said to be {\em
$P_l$-compatible} with a given cluster configuration if the action of that
cluster configuration on the connectivity state produces a final state that
differs from the initial one just by a permutation $P_l$ of the black points.
This generalises the notion of compatibility used in Sec.~\ref{sec:charKl}
to take into account the permutations of black points.

Let us first count the number of standard connectivities $|v_{l,i}\rangle$
which are $P_l$-compatible with a cluster configuration contributing to
$Z_{j,n_1,P}$. For $n_1=1$, $S_{n_1}$ contains only the identity element ${\rm
Id}$, and so the results of Sec.~\ref{sec:charKl} apply: the $Z_{j,1}$
contribute only to $K_{l,{\rm Id}}$. We consider next a configuration
contributing to $Z_{j,n_1,P}$ with $n_1>1$. The $|v_{l,i}\rangle$ which are
$P_l$-compatible with this configuration satisfy the same three rules as given
in Sec.~\ref{sec:charKl} for the case $n_1>1$, with the slight modification of
rule 3 that the black points must be attributed in such a
way that {\em the final state differs from the initial one by a permutation
$P_l$}. 

This modification makes the attribution of black points considerably more
involved than was the case in Sec.~\ref{sec:charKl}. 
First note that not all the $P_l$ are admissible.
To be precise, the cycle decomposition of the allowed
permutations can only contain 
$P$, as $P$ is the permutation 
between the branches realised by a single NTC. 
Therefore the admissible permutations contain
only $P$ and are such that $l=d_in_1$, denoting by $d_i$ the number of times $P$ is contained.
We note $(d_i,n_1)$ the corresponding classes
of permutations
and $K_{(d_i,n_1)}$ the corresponding $K$, see Eq.~(\ref{defKlCtor}).
Note that the number of classes of admissible permutations at a given level $l$ is
equal to the number of integers $d_i$ dividing $l$, i.e. $q(l)$.
Furthermore, inside these classes, not all permutations are admissible.
Indeed, the entanglement of
the NTC imply the entanglement of the structure of the allowed
permutations. We deduce from all this rules that, as announced, the admissible
permutations at level $l$ are simply the elements of the cyclic group $C_l$.

\begin{figure}
  \centering
  \includegraphics[width=300pt]{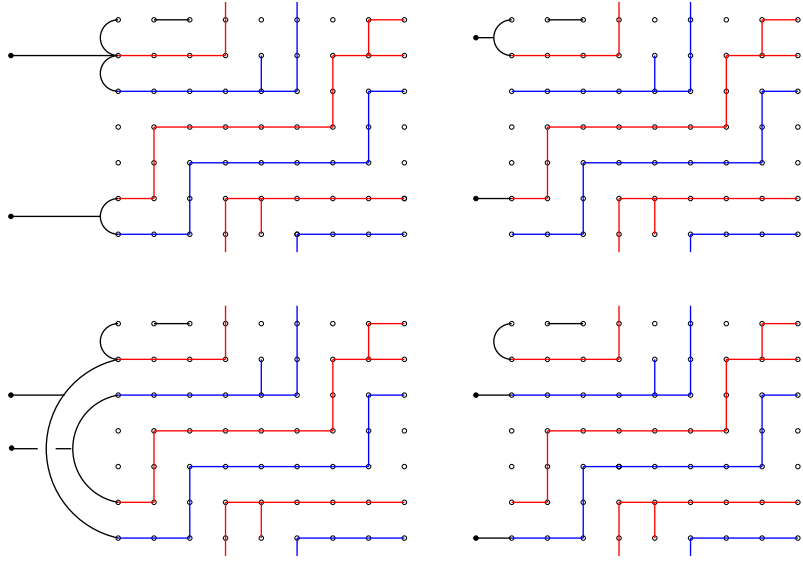}
  \caption{Standard connectivity states at level $l=2$ which are
  $(12)$-compatible with a given cluster configuration contributing to
  $Z_{2,2}$. The action of the cluster configuration on these connectivity
  states permutes the positions of the two black points.}
  \label{fig4}
\end{figure}

Let us now consider the decomposition of $K_{l,P_l}$, which is depicted in Fig.~\ref{fig4}, $P_l$
being an authorized permutation different from identity and containing $d_i$ times the permutation
$P$ of length $n_1$. Then, only the $Z_{j,n_1,P}$, with $j\geq d_i$, contribute to the decomposition 
of $K_{l,P_l}$. We find that 
the number of $|v_{l,i}\rangle$ which are $P_l$-compatible
with a given clusters configuration of $Z_{j,n_1,P}$ is ${2j \choose j-d_i}$.%
\footnote{Note that ${2j \choose j-d_i}$ is simply $n_{\rm tor}(j,d_i)$ for $d_i\geq 2$ but
is different for $d_i=1$, see Eq.~(\ref{defntor}).}
Therefore we have:
\begin{equation}
K_{l,P_l}=\sum_{j=d_i}^{\left \lfloor \frac{L}{n_1} \right \rfloor} {2j
\choose j-d_i} \frac{Z_{j,n_1,P}}{Q^j} \; .
\end{equation}
{}From this we infer the decomposition of $K_{(d_i,n_1)}$:
\begin{equation}
K_{(d_i,n_1)}=\sum_{j=d_i}^{\left \lfloor \frac{L}{n_1} \right \rfloor} {2j
\choose j-d_i} \frac{Z_{j,n_1}}{Q^j} \; .
\label{KnPn1}
\end{equation}
We will use the decomposition of $K_{(d_i,n_1)}$ in the following as 
it is simplier to work with $Z_{j,n_1}$ than with $Z_{j,n_1,P}$
(but one could consider the $Z_{j,n_1,P}$ too).

It remains to study the special case of $P_l = {\rm
Id}$. This is in fact trivial. Indeed, in that case, the value of $n_1$ in
$Z_{j,n_1}$ is no longer fixed, and one must sum over all possible values of
$n_1$, taking into account that the case of $n_1=1$ is particular. 
Since $K_{l,{\rm Id}}=\frac{K_l}{l}$, one obtains simply
Eqs.~(\ref{expKltor})--(\ref{expK0tor}) of Sec.~\ref{sec:charKl} up to a
global factor.

\subsection{Decomposition of $Z_j$ over the $K_{l,P_l}$}
\label{sec:expZj_KlCl}

To obtain the decomposition of $Z_{j,n_1}$ in terms of the $K_{l,P_l}$, 
we invert Eq.~(\ref{KnPn1}) for varying $d_i$ and
fixed $n_1>1$ and we obtain:
\begin{equation}
 Z_{j,n_1}=Q^j \sum_{d_i=j}^{\left \lfloor \frac{L}{n_1} \right \rfloor}
 (-1)^{d_i-j} \frac{2d_i}{d_i+j} {d_i+j \choose d_i-j}
 K_{(d_i,n_1)} \qquad \mbox{for }n_1>1 \; .
\label{expZjn1}
\end{equation}
Since the coefficients in this sum do not depend on $n_1$ (provided that
$n_1>1$), we can sum this relation over $n_1$ and write it as
\begin{equation}
 Z_{j,n_1>1}=Q^j \sum_{d_i=j}^{\left \lfloor \frac{L}{2} \right\rfloor}
 (-1)^{d_i-j} \frac{2d_i}{d_i+j} {d_i+j \choose d_i-j}
 K_{(d_i,n_1>1)}
\label{expZjn1>1}
\end{equation}
where we recall the notations $Z_{j,n_1>1}=\sum_{n_1=2}^L Z_{j,n_1}$
and $K_{(d_i,n_1>1)}=\sum_{n_1=2}^L K_{(d_i,n_1)}$, corresponding to
permutations consisting of $d_i$ cycles of the same length $>1$.

Consider next the case $n_1=1$. For $j\geq 2$ one has simply
\begin{equation}
Z_{j,1}=\sum_{l=j}^{L} \frac{b^{(l)}_{j}}{l} K_l \;,
\label{expZj12}
\end{equation}
recalling Eq.~(\ref{expZj1i}) and the fact that for $l\ge 2$ the
$Z_{j,n_1>1}$ do not appear in the decomposition of
$K_l$. However, according to Eqs.~(\ref{expK1tor})--(\ref{expK0tor}),
the $Z_{j,n_1>1}$ do appear for
$l=0$ and $l=1$, and one obtains
\begin{equation}
 Z_{1,1} = \sum_{l=1}^L \frac{b^{(l)}_{1}}{l} K_l 
 -\frac{Q}{2}\sum_{j=1}^{\left \lfloor \frac{L}{2} \right\rfloor}
 {2j \choose j} \frac{Z_{j,n_1>1}}{Q^j} \; .
\label{expZ11i}
\end{equation}
Inserting the decomposition (\ref{expZjn1>1}) of
$Z_{j,n_1>1}$ into Eq.~(\ref{expZ11i}) one obtains the decomposition of
$Z_{1,1}$ over $K_{l}$ and $K_{(d_i,n_1)}$:
\begin{equation}
Z_{1,1}=\sum_{l=1}^L \frac{b^{(l)}_1}{l} K_{l}+ Q \sum_{d_i=1}^{\left \lfloor \frac{L}{2} \right\rfloor} (-1)^{d_i} 
K_{(d_i,n_1>1)} \; .
\label{expZ11}
\end{equation}
We proceed in the same fashion for the decomposition of $Z_0 \equiv Z_{0,1}$,
finding
\begin{equation}
 Z_{0}=\sum_{l=0}^L \frac{b^{(l)}_{0}}{l} K_l - \frac{1}{2} \sum_{j=1}^{\left \lfloor \frac{L}{2} \right\rfloor}
 {2j \choose j} \frac{Z_{j,n_1>1}}{Q^j} \; .
\label{expZ0i}
\end{equation}
Upon insertion of the decomposition (\ref{expZjn1>1}) of $Z_{j,n_1>1}$, one
arrives at
\begin{equation}
 Z_{0}=\sum_{l=0}^L \frac{b^{(l)}_{0}}{l} K_l + 
 \sum_{d_i=1}^{\left \lfloor \frac{L}{2} \right\rfloor}  (-1)^{d_i} 
 K_{(d_i,n_1>1)} \; .
\label{expZ0}
\end{equation}

Since $Z_j=Z_{j,1}+Z_{j,n_1>1}$, we conclude from
Eqs.~(\ref{expZj12})--(\ref{expZjn1>1}) and from Eq.~(\ref{defbltil}) that, for any $j$,
\begin{equation}
 Z_j=\sum_{l=j}^L \frac{b^{(l)}_j}{l} K_{l}
 + \sum_{d_i=j}^{\left \lfloor \frac{L}{2} \right\rfloor} \tilde{b}^{(d_i)}_{j} K_{(d_i,n_1>1)} \; .
 \label{expZj}
\end{equation}

The decomposition of $Z \equiv \sum_{0 \leq j \leq L} Z_j$ is therefore
\begin{equation}
 Z=\sum_{l=0}^L \frac{b^{(l)}}{l} K_{l} +
 \sum_{d_i=1}^{\left \lfloor \frac{L}{2} \right\rfloor} \tilde{b}^{(d_i)} K_{(d_i,n_1>1)} \; .
\label{expZ}
\end{equation}

\section{Amplitudes of the eigenvalues}
\label{sec4}

\subsection{Decomposition of $Z$ over the $K_{l,D}$}

The culmination of the preceeding section was the decomposition (\ref{expZj})
of $Z_j$ in terms of $K_{l,P_l}$ (as $K_{(d_i,n_1)}$ is the sum of
the $K_{l,P_l}$ with $P_l$ being an element of $C_l$ belonging to the class $(d_i,n_1)$).
However, it is the $K_{l,D}$ which are
directly related to the eigenvalues of the transfer matrix ${\rm T}$. For that
reason, we now use the relation (\ref{KlCfD}) between the $K_{l,P_l}$ and
the $K_{l,D_k}$ to obtain the decomposition of $Z_j$ in terms of $K_{l,D_k}$. The
result is:
\begin{equation}
 Z_j=\sum_{l,D_k} b^{(l,D_k)}_j K_{l,D_k}
\label{expZj2}
\end{equation}
where the coefficients $b^{(l,D_k)}_j$ are given by
\begin{equation}
 b^{(l,D_k)}_j=\frac{b^{(l)}_j}{l} +
 \sum_{(d_i<l) | l} \frac{\tilde{b}^{(d_i)}_j}{l} \;
 \sum _{P_l \in \left(d_i,\frac{l}{d_i}\right)} \bar{\chi}_{D_k}(P_l) \; .
\label{defblDj}
\end{equation}
Indeed, $K_{l}=\sum_{D_k} K_{l,D_k}$, and since $K_{(d_i,n_1)}$ corresponds to the
level $l=d_i n_1$, we have $K_{(d_i,n_1)}=\sum_{D_k \in C_{d_in_1}}
\frac{\bar{\chi}_{D_k}((d_i,n_1))}{l} K_{d_i n_1,D_k}$. (Recall that $(d_i,n_1)$ is the
class of permutations consisting of $d_i$ cycles of the same length $n_1=\frac{l}{d_i}$.) As
explained in Sec.~\ref{sec:coef_bl}, the $b^{(l,D_k)}_j$ are not simply equal to
$\frac{b^{(l)}_j}{l}$ because of the $n_1>1$ terms. Using
Eq.~(\ref{relcDC}) we find that they nevertheless obey the following
relation
\begin{equation}
\sum_{D_k \in C_l} b^{(l,D_k)}_j=b^{(l)}_j \; .
\label{relblDj}
\end{equation}
But from Eq.~(\ref{defblDj}) the $b^{(l,D_k)}_j$ with $l<2j$ are trivial, i.e.,
equal to $\frac{b^{(l)}_j}{l}$ independently of $D$. This could have been shown
directly by considering the decomposition (\ref{expKltor}) of $K_l$. 

The decomposition of $Z$ over $K_{l,D_k}$ is obviously given by
\begin{equation}
Z=\sum_{l,D_k} b^{(l,D_k)} K_{l,D_k}
\label{expZKlD}
\end{equation}
where
\begin{equation}
b^{(l,D_k)}=\sum_{j=1}^l b^{(l,D_k)}_j \; ,
\end{equation}
i.e.
\begin{equation}
b^{(l,D_k)}=\frac{b^{(l)}}{l} +
 \sum_{(d_i<l) | l} \frac{\tilde{b}^{(d_i)}}{l} \;
 \sum _{P_l \in \left(d_i,\frac{l}{d_i}\right)} \bar{\chi}_{D_k}(P_l) \; .
\label{defblD}
\end{equation}
This is the central result of our article: we have obtained a rather simple
expression of the amplitudes $b^{(l,D)}$ in terms of the characters of
the irrep $D$.
A priori, for a given level $l$, there should be $l$
distinct amplitudes $b^{(l,D)}$ because $C_l$ has $l$ distinct irreps $D$.
However, because of the fact that two different permutations in the same
class $\left( d_i,\frac{l}{d_i}\right)$ correspond to the same coefficient $b^{(d_i)}$, there are
less distinct amplitudes: some $b^{(l,D)}$ are the same. Indeed, the 
Eq.~(\ref{defblD}) giving the amplitudes of the eigenvalues contains
generalized Ramanujan's sum, so using the subsection~\ref{Clprop}, the
$D_k$ whose $k$ are in the same $A_d$ correspond to the same amplitude
$b^{(l,D_d)}$. For example, at level $6$, there are only four distinct
amplitudes: $b^{(6,D_1)}$, $b^{(6,D_2)}$, $b^{(6,D_3)}$ and $b^{(6,D_6)}$,
since we have $b^{(6,D_1)}=b^{(6,D_5)}$ and $b^{(6,D_2)}=b^{(6,D_4)}$.

An important consequence of the expression of the $b^{(l,D_k)}$ is that
they satisfy
\begin{equation}
\sum_{D_k \in C_l} b^{(l,D_k)}=b^{(l)} \; ,
\label{relblD2}
\end{equation}
i.e., the sum of the $l$ (not necessarily distinct) amplitudes $b^{(l,D_k)}$
at level $l$ is equal to $b^{(l)}$. This has been previously noted by Chang
and Shrock~\cite{shrock}, except that they stated it was the sum of $l!$
amplitudes, not $l$, as they did not notice that only permutations in
the cyclic group $C_l$ were admissible.

Note also that for $l\geq 2$, Eq.~(\ref{defblD}) can be written more simply as:
\begin{equation}
b^{(l,D_k)}=
 \sum_{d_i | l} \frac{\tilde{b}^{(d_i)}}{l} \;
 \sum _{P_l \in \left(d_i,\frac{l}{d_i}\right)} \bar{\chi}_{D_k}(P_l) \; ,
\label{def2blD}
\end{equation}
since $b^{(l)}=\tilde{b}^{(l)}$ for $l\geq 2$. We now restrict ourselves to this case, as
the amplitudes at levels $0$ and $1$ are simply $b^{(0)}=1$ and $b^{(1)}=Q-1$.

\subsection{Compact formula for the amplitudes}
\label{sec:amplitudes}

We now calculate the Ramanujan's sums appearing in Eq.~(\ref{def2blD}). Using
Eq.~(\ref{passage}), we obtain:
\begin{equation}
b^{(l,m)}=
 \sum_{d_i | l} \frac{\mu \left(\frac{m}{m\wedge d_i}  \right)\phi \left( \frac{l}{d_i} \right)}{l\phi\left(\frac{m}{m\wedge d_i}  \right)} \;
 \tilde{b}^{(d_i)}
\end{equation}
Remember that $m$ is given by $\frac{l}{d}$ for $k$ in the set $A_{d}$, and so is an integer divisor of $l$.
Using the expression of the $\tilde{b}^{(d_i)}$ given in Eq.~(\ref{def2bltil}), we finally recover
the formula (\ref{eqsal}) of Read and Saleur. In particular, the term $(-1)^l(Q-1)$ in the definition~(\ref{defbltil}) of 
$\tilde{b}^{(l)}$ corresponds to degenerate cluster configurations.

Note that the number of different amplitudes at level $l$ is simply equal to the number of integer divisors
of $l$. In particular, when $l$ is prime, there are only two different amplitudes: $b^{(l,1)}$ which
corresponds to $b^{(l,D_l)}$ ($D_l$ is the identity representation) and $b^{(l,l)}$ which corresponds
to the $l-1$ other $b^{(l,D_k)}$ (as they are all equal). Using that $b^{(1)}=-1$, we find:
\begin{eqnarray}
b^{(l,1)}&=&\frac{b^{(l)}-l+1}{l}  \\
b^{(l,l)}&=&\frac{b^{(l)}+1}{l}
\end{eqnarray}
This could have been simply directly showed using Eq.~(\ref{def2blD}). Indeed, for $l$ prime, $C_l$ contains
${\rm Id}$ and $l-1$ cycles of length $l$. As $b^{(1)}=-1$, we deduce that $b^{(l,1)}=\frac{b^{(l)}-l+1}{l}$.
For $b^{(l,l)}$, one needs just use that $\sum_{k=1}^{l-1} \exp \left( \frac{i2\pi k}{l}\right) =-1$.

\section{Conclusion}

To summarise, we have generalised the combinatorial approach developed in
Ref.~\cite{cyclic} for cyclic boundary conditions to the case of toroidal
boundary conditions. In particular, we have obtained the decomposition of the
partition function for the Potts model on finite tori in terms of the
generalised characters $K_{l,D}$. We proved that the formula (\ref{eqsal}) of Read and
Saleur is valid for any finite lattice, and for any inhomogeneous choice of the coupling constants. Furthermore, our physical
interpretation of this formula is new and is based on the cyclic group
$C_l$.

The eigenvalue amplitudes are instrumental in determining the physics of the
Potts model, in particular in the antiferromagnetic regime~\cite{hs,af}.
Generically, this regime belongs to a so-called Berker-Kadanoff (BK) phase in
which the temperature variable is irrelevant in the renormalisation group
sense, and whose properties can be obtained by analytic continuation of the
well-known ferromagnetic phase transition~\cite{hs}. Due to the
Beraha-Kahane-Weiss (BKW) theorem~\cite{bkw}, partition function zeros
accumulate at the values of $Q$ where either the amplitude of the dominant
eigenvalue vanishes, or where the two dominant eigenvalues become equimodular.
When this happens, the BK phase disappears, and the system undergoes a phase
transition with control parameter $Q$. Determining analytically the eigenvalue
amplitudes is thus directly relevant for the first of the hypotheses in the
BKW theorem.

For the cyclic geometry, the amplitudes are very simple, and the real values of $Q$
satisfying the hypothesis of the BKW theorem are simply the so-called Beraha
numbers, $Q=B_n=(2 \cos(\pi/n))^2$ with $n=2,3,\ldots$, independently of the
width $L$. For the toroidal case, the formula is more complicated, and there can be 
degeneracies of eigenvalues between different levels which depend on the width $L$ of the lattice, as shown by Chang and Shrock~\cite{shrock}. The role of the Beraha numbers will therefore
be considered in a future work.

\vspace{0.1cm}

\noindent
{\bf Acknowledgments.}

The authors are grateful to H.~Saleur, J.-B.~Zuber and P.~Zinn-Justin for some
useful discussions. We also thank J.~Salas for collaboration on closely
related projects.

\newpage
\thispagestyle{empty}
\strut

\end{document}